\documentclass[10pt,a4paper,onecolumn]{article}
\usepackage{marginnote}
\usepackage{graphicx}
\usepackage{xcolor}
\usepackage{authblk,etoolbox}
\usepackage{titlesec}
\usepackage{calc}
\usepackage{tikz}
\usepackage{hyperref}
\hypersetup{colorlinks,breaklinks,
            urlcolor=[rgb]{0.0, 0.5, 1.0},
            linkcolor=[rgb]{0.0, 0.5, 1.0}}
\usepackage{caption}
\usepackage{tcolorbox}
\usepackage{amssymb,amsmath}
\usepackage{ifxetex,ifluatex}
\usepackage{seqsplit}
\usepackage{xstring}

\usepackage{float}
\let\origfigure\figure
\let\endorigfigure\endfigure
\renewenvironment{figure}[1][2] {
    \expandafter\origfigure\expandafter[H]
} {
    \endorigfigure
}

\usepackage{fixltx2e} 
\usepackage[
  backend=biber,
]{biblatex}
\bibliography{paper.bib}

\let\textttOrig=\texttt
\def\texttt#1{\expandafter\textttOrig{\seqsplit{#1}}}
\renewcommand{\seqinsert}{\ifmmode
  \allowbreak
  \else\penalty6000\hspace{0pt plus 0.02em}\fi}

\makeatletter
\let\href@Orig=\href
\def\href@Urllike#1#2{\href@Orig{#1}{\begingroup
    \def\Url@String{#2}\Url@FormatString
    \endgroup}}
\def\href@Notdoi#1#2{\def\tempa{#1}\def\tempb{#2}%
  \ifx\tempa\tempb\relax\href@Urllike{#1}{#2}\else
  \href@Orig{#1}{#2}\fi}
\def\href#1#2{%
  \IfBeginWith{#1}{https://doi.org}%
  {\href@Urllike{#1}{#2}}{\href@Notdoi{#1}{#2}}}
\makeatother

\usepackage[top=3.5cm, bottom=3cm, right=1.5cm, left=1.0cm,
            headheight=2.2cm, reversemp, includemp, marginparwidth=4.5cm]{geometry}

\titleformat{\section}
  {\normalfont\sffamily\Large\bfseries}
  {}{0pt}{}
\titleformat{\subsection}
  {\normalfont\sffamily\large\bfseries}
  {}{0pt}{}
\titleformat{\subsubsection}
  {\normalfont\sffamily\bfseries}
  {}{0pt}{}
\titleformat*{\paragraph}
  {\sffamily\normalsize}

\usepackage{fancyhdr}
\makeatletter
\let\ps@plain\ps@fancy
\fancyheadoffset[L]{4.5cm}
\fancyfootoffset[L]{4.5cm}

\definecolor{linky}{rgb}{0.0, 0.5, 1.0}

\newtcolorbox{repobox}
   {colback=red, colframe=red!75!black,
     boxrule=0.5pt, arc=2pt, left=6pt, right=6pt, top=3pt, bottom=3pt}

\newcommand{\ExternalLink}{%
   \tikz[x=1.2ex, y=1.2ex, baseline=-0.05ex]{%
       \begin{scope}[x=1ex, y=1ex]
           \clip (-0.1,-0.1)
               --++ (-0, 1.2)
               --++ (0.6, 0)
               --++ (0, -0.6)
               --++ (0.6, 0)
               --++ (0, -1);
           \path[draw,
               line width = 0.5,
               rounded corners=0.5]
               (0,0) rectangle (1,1);
       \end{scope}
       \path[draw, line width = 0.5] (0.5, 0.5)
           -- (1, 1);
       \path[draw, line width = 0.5] (0.6, 1)
           -- (1, 1) -- (1, 0.6);
       }
   }

\patchcmd{\@maketitle}{center}{flushleft}{}{}
\patchcmd{\@maketitle}{center}{flushleft}{}{}
\patchcmd{\@maketitle}{\LARGE}{\LARGE\sffamily}{}{}
\def\maketitle{{%
  
  \AB@maketitle}}
\makeatletter
\renewcommand\AB@affilsepx{ \protect\Affilfont}
\renewcommand\AB@affilnote[1]{{\bfseries #1}\hspace{3pt}}
\renewcommand{\affil}[2][]%
   {\newaffiltrue\let\AB@blk@and\AB@pand
      \if\relax#1\relax\def\AB@note{\AB@thenote}\else\def\AB@note{#1}%
        \setcounter{Maxaffil}{0}\fi
        \begingroup
        \let\href=\href@Orig
        \let\texttt=\textttOrig
        \let\protect\@unexpandable@protect
        \def\thanks{\protect\thanks}\def\footnote{\protect\footnote}%
        \@temptokena=\expandafter{\AB@authors}%
        {\def\\{\protect\\\protect\Affilfont}\xdef\AB@temp{#2}}%
         \xdef\AB@authors{\the\@temptokena\AB@las\AB@au@str
         \protect\\[\affilsep]\protect\Affilfont\AB@temp}%
         \gdef\AB@las{}\gdef\AB@au@str{}%
        {\def\\{, \ignorespaces}\xdef\AB@temp{#2}}%
        \@temptokena=\expandafter{\AB@affillist}%
        \xdef\AB@affillist{\the\@temptokena \AB@affilsep
          \AB@affilnote{\AB@note}\protect\Affilfont\AB@temp}%
      \endgroup
       \let\AB@affilsep\AB@affilsepx
}
\makeatother

\renewcommand\Affilfont{\sffamily\small\mdseries}
\ifnum 0\ifxetex 1\fi\ifluatex 1\fi=0 
  \usepackage[T1]{fontenc}
  \usepackage[utf8]{inputenc}

\else 
  \ifxetex
    \usepackage{mathspec}
  \else
    \usepackage{fontspec}
  \fi
  \defaultfontfeatures{Ligatures=TeX,Scale=MatchLowercase}

\fi
\IfFileExists{upquote.sty}{\usepackage{upquote}}{}
\IfFileExists{microtype.sty}{%
\usepackage{microtype}
\UseMicrotypeSet[protrusion]{basicmath} 
}{}

\usepackage{hyperref}
\hypersetup{unicode=true,
            pdftitle={c-lasso: a Python package for constrained sparse regression and classification},
            pdfborder={0 0 0},
            breaklinks=true}
\usepackage{color}
\usepackage{fancyvrb}

\DefineVerbatimEnvironment{Highlighting}{Verbatim}{commandchars=\\\{\}}
\newenvironment{Shaded}{}{}
\newcommand{\KeywordTok}[1]{\textcolor[rgb]{0.00,0.44,0.13}{\textbf{#1}}}
\newcommand{\DataTypeTok}[1]{\textcolor[rgb]{0.56,0.13,0.00}{#1}}
\newcommand{\DecValTok}[1]{\textcolor[rgb]{0.25,0.63,0.44}{#1}}

\newcommand{\FloatTok}[1]{\textcolor[rgb]{0.25,0.63,0.44}{#1}}

\newcommand{\SpecialCharTok}[1]{\textcolor[rgb]{0.25,0.44,0.63}{#1}}
\newcommand{\StringTok}[1]{\textcolor[rgb]{0.25,0.44,0.63}{#1}}

\newcommand{\ImportTok}[1]{#1}
\newcommand{\CommentTok}[1]{\textcolor[rgb]{0.38,0.63,0.69}{\textit{#1}}}

\newcommand{\OtherTok}[1]{\textcolor[rgb]{0.00,0.44,0.13}{#1}}

\newcommand{\VariableTok}[1]{\textcolor[rgb]{0.10,0.09,0.49}{#1}}

\newcommand{\OperatorTok}[1]{\textcolor[rgb]{0.40,0.40,0.40}{#1}}
\newcommand{\BuiltInTok}[1]{#1}

\newcommand{\NormalTok}[1]{#1}
\usepackage{longtable,booktabs}

\let\addcontentslineOrig=\addcontentsline
\def\addcontentsline#1#2#3{\bgroup
  \let\texttt=\textttOrig\addcontentslineOrig{#1}{#2}{#3}\egroup}
\let\markbothOrig\markboth
\def\markboth#1#2{\bgroup
  \let\texttt=\textttOrig\markbothOrig{#1}{#2}\egroup}
\let\markrightOrig\markright
\def\markright#1{\bgroup
  \let\texttt=\textttOrig\markrightOrig{#1}\egroup}

\usepackage{graphicx,grffile}
\makeatletter
\def\maxwidth{\ifdim\Gin@nat@width>\linewidth\linewidth\else\Gin@nat@width\fi}
\def\maxheight{\ifdim\Gin@nat@height>\textheight\textheight\else\Gin@nat@height\fi}
\makeatother
\setkeys{Gin}{width=\maxwidth,height=\maxheight,keepaspectratio}
\IfFileExists{parskip.sty}{%
\usepackage{parskip}
}{
\setlength{\parindent}{0pt}
\setlength{\parskip}{6pt plus 2pt minus 1pt}
}
\ifx\paragraph\undefined\else
\let\oldparagraph\paragraph
\renewcommand{\paragraph}[1]{\oldparagraph{#1}\mbox{}}
\fi
\ifx\subparagraph\undefined\else
\let\oldsubparagraph\subparagraph
\renewcommand{\subparagraph}[1]{\oldsubparagraph{#1}\mbox{}}
\fi

\title{c-lasso: a Python package for constrained sparse regression and classification}

        \author[1]{L\'eo Simpson}
          \author[2]{Patrick L. Combettes}
          \author[3,4,5]{Christian L. M\"uller}
    
      \affil[1]{Technische Universität M\"unchen}
      \affil[2]{Department of Mathematics, North Carolina State University, Raleigh}
      \affil[3]{Center for Computational Mathematics, Flatiron Institute, New York}
      \affil[4]{Institute of Computational Biology, Helmholtz Zentrum M\"unchen}
      \affil[5]{Department of Statistics, Ludwig-Maximilians-Universität M\"unchen}
  \date{\vspace{-7ex}}

\begin{document}
\maketitle

\marginpar{
  \sffamily\small

  {\bfseries DOI:} \href{https://doi.org/}{\color{linky}{}}

  \vspace{2mm}

  {\bfseries Software}
  \begin{itemize}
    \setlength\itemsep{0em}
    \item \href{}{\color{linky}{Review}} \ExternalLink
    \item \href{https://github.com/Leo-Simpson/c-lasso}{\color{linky}{Repository}} \ExternalLink
    \item \href{}{\color{linky}{Archive}} \ExternalLink
  \end{itemize}

  \vspace{2mm}

  {\bfseries Submitted:} November 2020\\
  {\bfseries Published:} 

  \vspace{2mm}
  {\bfseries License}\\
  Authors of papers retain copyright and release the work under a Creative Commons Attribution 4.0 International License (\href{https://creativecommons.org/licenses/by/4.0/}{\color{linky}{CC BY 4.0}}).
}

\section{Summary}\label{summary}

We introduce \texttt{c-lasso}, a Python package that enables sparse and
robust linear regression and classification with linear equality
constraints. The underlying statistical forward model is assumed to be
of the following form:

\[
y = X \beta + \sigma \epsilon \qquad \textrm{subject to} \qquad C\beta=0
\]

Here, \(X \in \mathbb{R}^{n\times d}\) is a given design matrix and the
vector \(y \in \mathbb{R}^{n}\) is a continuous or binary response
vector. The matrix \(C\) is a general constraint matrix. The vector
\(\beta \in \mathbb{R}^{d}\) contains the unknown coefficients and
\(\sigma\) an unknown scale. Prominent use cases are (sparse)
log-contrast regression with compositional data \(X\), requiring the
constraint \(1_d^T \beta = 0\) (Aitchion and Bacon-Shone 1984) and the
Generalized Lasso which is a \emph{special case} of the described
problem (see, e.g, (James, Paulson, and Rusmevichientong 2020), Example
3). The \texttt{c-lasso} package provides estimators for
inferring unknown coefficients and scale (i.e., perspective M-estimators
(Combettes and M\"uller 2020a)) of the form

\[
    \min_{\beta \in \mathbb{R}^d, \sigma \in \mathbb{R}_{0}} f\left(X\beta - y,{\sigma} \right) + \lambda \left\lVert \beta\right\rVert_1 \qquad \textrm{subject to} \qquad  C\beta = 0
\]

for several convex loss functions \(f(\cdot,\cdot)\). This includes the
constrained Lasso, the constrained scaled Lasso, and sparse Huber
M-estimators with linear equality constraints.

\section{Statement of need}\label{statement-of-need}

Currently, there is no Python package available that can solve these
ubiquitous statistical estimation problems in a fast and efficient
manner. \texttt{c-lasso} provides algorithmic strategies, including path
and proximal splitting algorithms, to solve the underlying convex
optimization problems with provable convergence guarantees. The
\texttt{c-lasso} package is intended to fill the gap between popular
Python tools such as
\href{https://scikit-learn.org/stable/}{\texttt{scikit-learn}} which
cannot solve these constrained problems and general-purpose optimization
solvers such as \href{https://www.cvxpy.org}{\texttt{cvxpy}} that do not
scale well for these problems and/or are inaccurate. \texttt{c-lasso}
can solve the estimation problems at a single regularization level,
across an entire regularization path, and includes three model selection
strategies for determining the regularization parameter: a
theoretically-derived fixed regularization, k-fold cross-validation, and
stability selection. We show several use cases of the package, including
an application of sparse log-contrast regression tasks for compositional
microbiome data, and highlight the seamless integration into \texttt{R}
via \href{https://rstudio.github.io/reticulate/}{\texttt{reticulate}}.

\section{Functionalities}\label{functionalities}

\subsection{Installation and problem
instantiation}\label{gettingstarted}

\texttt{c-lasso} is available on pip and can be installed in the shell
using

\begin{verbatim}
pip install c-lasso
\end{verbatim}

The central object in the \texttt{c-lasso} package is the instantiation
of a \texttt{c-lasso} problem.

\begin{Shaded}
\begin{Highlighting}[]
\CommentTok{# Import the main class of the package}
\ImportTok{from}\NormalTok{ classo }\ImportTok{import}\NormalTok{ classo_problem}

\CommentTok{# Define a c-lasso problem instance with default setting, }
\CommentTok{# given data X, y, and constraints C.}
\NormalTok{problem  }\OperatorTok{=}\NormalTok{ classo_problem(X,y,C)}
\end{Highlighting}
\end{Shaded}

We next describe what type of problem instances are available and how to
solve them.

\hypertarget{formulations}{\subsection{Statistical problem
formulations}\label{formulations}}

Depending on the type of and the prior assumptions on the data, the
noise \(\epsilon\), and the model parameters, \texttt{c-lasso} allows
for different estimation problem formulations. More specifically, the
package can solve the following four regression-type and two
classification-type formulations:

\hypertarget{R1}{\subsubsection{\texorpdfstring{\emph{R1} Standard
constrained Lasso
regression:}{R1 Standard constrained Lasso regression:}}\label{R1}}

\[
    \min_{\beta \in \mathbb{R}^d} \left\lVert X\beta - y \right\rVert^2 + \lambda \left\lVert \beta\right\rVert_1 \qquad \textrm{subject to} \qquad  C\beta = 0
\]

This is the standard Lasso problem with linear equality constraints on
the \(\beta\) vector. The objective function combines Least-Squares (LS)
for model fitting with the \(L_1\)-norm penalty for sparsity.

\begin{Shaded}
\begin{Highlighting}[]
\CommentTok{# Formulation R1}
\NormalTok{problem.formulation.huber }\OperatorTok{=} \VariableTok{False}
\NormalTok{problem.formulation.concomitant }\OperatorTok{=} \VariableTok{False}
\NormalTok{problem.formulation.classification }\OperatorTok{=} \VariableTok{False}
\end{Highlighting}
\end{Shaded}

\hypertarget{R2}{\subsubsection{\texorpdfstring{\emph{R2} Contrained
sparse Huber
regression:}{R2 Contrained sparse Huber regression:}}\label{R2}}

\[
    \min_{\beta \in \mathbb{R}^d} h_{\rho} (X\beta - y) + \lambda \left\lVert \beta\right\rVert_1 \qquad \textrm{subject to} \qquad  C\beta = 0
\]

This regression problem uses the
\href{https://en.wikipedia.org/wiki/Huber_loss}{Huber loss} \(h_{\rho}\)
as objective function for robust model fitting with an \(L_1\) penalty
and linear equality constraints on the \(\beta\) vector. The default
parameter \(\rho\) is set to \(1.345\) (Huber 1981).

\begin{Shaded}
\begin{Highlighting}[]
\CommentTok{# Formulation R2}
\NormalTok{problem.formulation.huber }\OperatorTok{=} \VariableTok{True}
\NormalTok{problem.formulation.concomitant }\OperatorTok{=} \VariableTok{False}
\NormalTok{problem.formulation.classification }\OperatorTok{=} \VariableTok{False}
\end{Highlighting}
\end{Shaded}

\hypertarget{R3}{\subsubsection{\texorpdfstring{\emph{R3} Contrained
scaled Lasso
regression:}{R3 Contrained scaled Lasso regression:}}\label{R3}}

\[
    \min_{\beta \in \mathbb{R}^d, \sigma \in \mathbb{R}_{0}} \frac{\left\lVert X\beta - y \right\rVert^2}{\sigma} + \frac{n}{2} \sigma + \lambda \left\lVert \beta\right\rVert_1 \qquad \textrm{subject to} \qquad  C\beta = 0
\]

This formulation is the default problem formulation in \texttt{c-lasso}.
It is similar to \protect\hyperlink{R1}{\emph{R1}} but allows for joint
estimation of the (constrained) \(\beta\) vector and the standard
deviation \(\sigma\) in a concomitant fashion (Combettes and M\"uller
2020a; Combettes and M\"uller 2020b).

\begin{Shaded}
\begin{Highlighting}[]
\CommentTok{# Formulation R3}
\NormalTok{problem.formulation.huber }\OperatorTok{=} \VariableTok{False}
\NormalTok{problem.formulation.concomitant }\OperatorTok{=} \VariableTok{True}
\NormalTok{problem.formulation.classification }\OperatorTok{=} \VariableTok{False}
\end{Highlighting}
\end{Shaded}

\hypertarget{R4}{\subsubsection{\texorpdfstring{\emph{R4} Contrained
sparse Huber regression with concomitant scale
estimation:}{R4 Contrained sparse Huber regression with concomitant scale estimation:}}\label{R4}}

\[
    \min_{\beta \in \mathbb{R}^d, \sigma \in  \mathbb{R}_{0}} \left( h_{\rho} \left( \frac{X\beta - y}{\sigma} \right) + n \right) \sigma + \lambda \left\lVert \beta\right\rVert_1 \qquad \textrm{subject to} \qquad  C\beta = 0
\]

This formulation combines \protect\hyperlink{R2}{\emph{R2}} and
\protect\hyperlink{R3}{\emph{R3}} allowing robust joint estimation of
the (constrained) \(\beta\) vector and the scale \(\sigma\) in a
concomitant fashion (Combettes and M\"uller 2020a; Combettes and M\"uller
2020b).

\begin{Shaded}
\begin{Highlighting}[]
\CommentTok{# Formulation R4}
\NormalTok{problem.formulation.huber }\OperatorTok{=} \VariableTok{True}
\NormalTok{problem.formulation.concomitant }\OperatorTok{=} \VariableTok{True}
\NormalTok{problem.formulation.classification }\OperatorTok{=} \VariableTok{False}
\end{Highlighting}
\end{Shaded}

\hypertarget{C1}{\subsubsection{\texorpdfstring{\emph{C1} Contrained
sparse classification with Square Hinge
loss:}{C1 Contrained sparse classification with Square Hinge loss:}}\label{C1}}

\[
    \min_{\beta \in \mathbb{R}^d} \sum_{i=1}^n l(y_i x_i^\top\beta) + \lambda \left\lVert \beta\right\rVert_1 \qquad \textrm{subject to} \qquad  C\beta = 0
\]

where \(x_i\) denotes the \(i^{th}\) row of \(X\), \(y_i \in \{-1,1\}\),
and \(l(\cdot)\) is defined for \(r \in \mathbb{R}\) as:

\[
l(r) = \begin{cases} (1-r)^2 & if \quad r \leq 1 \\ 0 &if \quad r \geq 1 \end{cases}
\]

This formulation is similar to \protect\hyperlink{R1}{\emph{R1}} but
adapted for classification tasks using the Square Hinge loss with
(constrained) sparse \(\beta\) vector estimation (Lee and Lin 2013).

\begin{Shaded}
\begin{Highlighting}[]
\CommentTok{# Formulation C1}
\NormalTok{problem.formulation.huber }\OperatorTok{=} \VariableTok{False}
\NormalTok{problem.formulation.concomitant }\OperatorTok{=} \VariableTok{False}
\NormalTok{problem.formulation.classification }\OperatorTok{=} \VariableTok{True}
\end{Highlighting}
\end{Shaded}

\hypertarget{C2}{\subsubsection{\texorpdfstring{\emph{C2} Contrained
sparse classification with Huberized Square Hinge
loss:}{C2 Contrained sparse classification with Huberized Square Hinge loss:}}\label{C2}}

\[
    \min_{\beta \in \mathbb{R}^d}  \sum_{i=1}^n  l_{\rho}(y_i x_i^\top\beta) + \lambda \left\lVert \beta\right\rVert_1 \qquad \textrm{subject to} \qquad  C\beta = 0 \,.
\]

This formulation is similar to \protect\hyperlink{C1}{\emph{C1}} but
uses the Huberized Square Hinge loss \(l_{\rho}\) for robust
classification with (constrained) sparse \(\beta\) vector estimation
(Rosset and Zhu 2007):

\[
l_{\rho}(r) = \begin{cases} (1-r)^2 &if \quad \rho \leq r \leq 1 \\ (1-\rho)(1+\rho-2r) & if \quad r \leq \rho \\ 0 &if \quad r \geq 1 \end{cases}
\]

This formulation can be selected in \texttt{c-lasso} as follows:

\begin{Shaded}
\begin{Highlighting}[]
\CommentTok{# Formulation C2}
\NormalTok{problem.formulation.huber }\OperatorTok{=} \VariableTok{True}
\NormalTok{problem.formulation.concomitant }\OperatorTok{=} \VariableTok{False}
\NormalTok{problem.formulation.classification }\OperatorTok{=} \VariableTok{True}
\end{Highlighting}
\end{Shaded}

\subsection{Optimization schemes}\label{method}

The problem formulations \emph{R1}-\emph{C2} require different
algorithmic strategies for efficiently solving the underlying
optimization problems. The \texttt{c-lasso} package implements four
published algorithms with provable convergence guarantees. The package
also includes novel algorithmic extensions to solve Huber-type problems
using the mean-shift formulation (Mishra and M\"uller 2019). The following
algorithmic schemes are implemented:

\begin{itemize}
\item
  Path algorithms (\emph{Path-Alg}): This algorithm follows the proposal
  in (Gaines, Kim, and Zhou 2018; Jeon et al. 2020) and uses the fact
  that the solution path along \(\lambda\) is piecewise-affine (Rosset
  and Zhu 2007). We also provide a novel efficient procedure that allows
  to derive the solution for the concomitant problem \emph{R3} along the
  path with little computational overhead.
\item
  Douglas-Rachford-type splitting method (\emph{DR}): This algorithm can
  solve all regression problems \emph{R1-R4}. It is based on
  Doulgas-Rachford splitting in a higher-dimensional product space and
  makes use of the proximity operators of the perspective of the LS
  objective (Combettes and M\"uller 2020a; Combettes and M\"uller 2020b).
  The Huber problem with concomitant scale \emph{R4} is reformulated as
  scaled Lasso problem with mean shift vector (Mishra and M\"uller 2019)
  and thus solved in (n + d) dimensions.
\item
  Projected primal-dual splitting method (\emph{P-PDS}): This algorithm
  is derived from (Briceño-Arias and López Rivera 2019) and belongs to
  the class of proximal splitting algorithms, extending the classical
  Forward-Backward (FB) (aka proximal gradient descent) algorithm to
  handle an additional linear equality constraint via projection. In the
  absence of a linear constraint, the method reduces to FB.
\item
  Projection-free primal-dual splitting method (\emph{PF-PDS}): This
  algorithm is a special case of an algorithm proposed in (Combettes and
  Pesquet 2012) (Eq. 4.5) and also belongs to the class of proximal
  splitting algorithms. The algorithm does not require projection
  operators which may be beneficial when C has a more complex structure.
  In the absence of a linear constraint, the method reduces to the
  Forward-Backward-Forward scheme.
\end{itemize}

The following table summarizes the available algorithms and their
recommended use for each problem:

\begin{longtable}[]{@{}lcccc@{}}
\toprule
\begin{minipage}[b]{0.02\columnwidth}\raggedright\strut
\strut
\end{minipage} & \begin{minipage}[b]{0.2\columnwidth}\centering\strut
\emph{Path-Alg}\strut
\end{minipage} & \begin{minipage}[b]{0.2\columnwidth}\centering\strut
\emph{DR}\strut
\end{minipage} & \begin{minipage}[b]{0.2\columnwidth}\centering\strut
\emph{P-PDS}\strut
\end{minipage} & \begin{minipage}[b]{0.2\columnwidth}\centering\strut
\emph{PF-PDS}\strut
\end{minipage}\tabularnewline
\midrule
\endhead
\begin{minipage}[t]{0.02\columnwidth}\raggedright\strut
\protect\hyperlink{R1}{\emph{R1}}\strut
\end{minipage} & \begin{minipage}[t]{0.2\columnwidth}\centering\strut
use for large \(\lambda\) and path computation\strut
\end{minipage} & \begin{minipage}[t]{0.2\columnwidth}\centering\strut
use for small \(\lambda\)\strut
\end{minipage} & \begin{minipage}[t]{0.2\columnwidth}\centering\strut
possible\strut
\end{minipage} & \begin{minipage}[t]{0.2\columnwidth}\centering\strut
use for complex constraints\strut
\end{minipage}\tabularnewline
\begin{minipage}[t]{0.02\columnwidth}\raggedright\strut
\protect\hyperlink{R2}{\emph{R2}}\strut
\end{minipage} & \begin{minipage}[t]{0.2\columnwidth}\centering\strut
use for large \(\lambda\) and path computation\strut
\end{minipage} & \begin{minipage}[t]{0.2\columnwidth}\centering\strut
use for small \(\lambda\)\strut
\end{minipage} & \begin{minipage}[t]{0.2\columnwidth}\centering\strut
possible\strut
\end{minipage} & \begin{minipage}[t]{0.2\columnwidth}\centering\strut
use for complex constraints\strut
\end{minipage}\tabularnewline
\begin{minipage}[t]{0.02\columnwidth}\raggedright\strut
\protect\hyperlink{R3}{\emph{R3}}\strut
\end{minipage} & \begin{minipage}[t]{0.2\columnwidth}\centering\strut
use for large \(\lambda\) and path computation\strut
\end{minipage} & \begin{minipage}[t]{0.2\columnwidth}\centering\strut
use for small \(\lambda\)\strut
\end{minipage} & \begin{minipage}[t]{0.2\columnwidth}\centering\strut
-\strut
\end{minipage} & \begin{minipage}[t]{0.2\columnwidth}\centering\strut
-\strut
\end{minipage}\tabularnewline
\begin{minipage}[t]{0.02\columnwidth}\raggedright\strut
\protect\hyperlink{R4}{\emph{R4}}\strut
\end{minipage} & \begin{minipage}[t]{0.2\columnwidth}\centering\strut
-\strut
\end{minipage} & \begin{minipage}[t]{0.2\columnwidth}\centering\strut
only option\strut
\end{minipage} & \begin{minipage}[t]{0.2\columnwidth}\centering\strut
-\strut
\end{minipage} & \begin{minipage}[t]{0.2\columnwidth}\centering\strut
-\strut
\end{minipage}\tabularnewline
\begin{minipage}[t]{0.02\columnwidth}\raggedright\strut
\protect\hyperlink{C1}{\emph{C1}}\strut
\end{minipage} & \begin{minipage}[t]{0.2\columnwidth}\centering\strut
only option\strut
\end{minipage} & \begin{minipage}[t]{0.2\columnwidth}\centering\strut
-\strut
\end{minipage} & \begin{minipage}[t]{0.2\columnwidth}\centering\strut
-\strut
\end{minipage} & \begin{minipage}[t]{0.2\columnwidth}\centering\strut
-\strut
\end{minipage}\tabularnewline
\begin{minipage}[t]{0.02\columnwidth}\raggedright\strut
\protect\hyperlink{C2}{\emph{C2}}\strut
\end{minipage} & \begin{minipage}[t]{0.2\columnwidth}\centering\strut
only option\strut
\end{minipage} & \begin{minipage}[t]{0.2\columnwidth}\centering\strut
-\strut
\end{minipage} & \begin{minipage}[t]{0.2\columnwidth}\centering\strut
-\strut
\end{minipage} & \begin{minipage}[t]{0.2\columnwidth}\centering\strut
-\strut
\end{minipage}\tabularnewline
\bottomrule
\end{longtable}

The following Python snippet shows how to select a specific algorithm:

\begin{Shaded}
\begin{Highlighting}[]
\NormalTok{problem.numerical_method }\OperatorTok{=} \StringTok{"Path-Alg"} 
\CommentTok{# Alternative options: "DR", "P-PDS", and "PF-PDS" }
\end{Highlighting}
\end{Shaded}

\hypertarget{model}{\subsection{Computation modes and model
selection}\label{model}}

The \texttt{c-lasso} package provides several computation modes and
model selection schemes for tuning the regularization parameter.

\begin{itemize}
\item
  \emph{Fixed Lambda}: This setting lets the user choose a fixed
  parameter \(\lambda\) or a proportion \(l \in [0,1]\) such that
  \(\lambda = l\times \lambda_{\max}\). The default value is a
  scale-dependent tuning parameter that has been derived in (Shi, Zhang,
  and Li 2016) and applied in (Combettes and M\"uller 2020b).
\item
  \emph{Path Computation}: This setting allows the computation of a
  solution path for \(\lambda\) parameters in an interval
  \([\lambda_{\min}, \lambda_{\max}]\). The solution path is computed
  via the \emph{Path-Alg} scheme or via warm-starts for other
  optimization schemes.
\end{itemize}

\begin{itemize}
\item
  \emph{Cross Validation}: This setting allows the selection of the
  regularization parameter \(\lambda\) via k-fold cross validation for
  \(\lambda \in [\lambda_{\min}, \lambda_{\max}]\). Both the Minimum
  Mean Squared Error (or Deviance) (MSE) and the ``One-Standard-Error
  rule'' (1SE) are available (Hastie, Tibshirani, and Friedman 2009).
\item
  \emph{Stability Selection}: This setting allows the selection of the
  \(\lambda\) via stability selection (Meinshausen and B\"uhlmann 2010;
  Lin et al. 2014; Combettes and M\"uller 2020b). Three modes are
  available: selection at a fixed \(\lambda\) (Combettes and M\"uller
  2020b), selection of the q \emph{first} variables entering the path
  (default setting), and of the q \emph{largest coefficients} (in
  absolute value) across the path (Meinshausen and B\"uhlmann 2010).
\end{itemize}

The Python syntax to use a specific computation mode and model selection
is exemplified below:

\begin{Shaded}
\begin{Highlighting}[]
\CommentTok{# Example how to perform ath computation and cross-validation:}
\NormalTok{problem.model_selection.LAMfixed }\OperatorTok{=} \VariableTok{False}
\NormalTok{problem.model_selection.PATH }\OperatorTok{=} \VariableTok{True}
\NormalTok{problem.model_selection.CV }\OperatorTok{=} \VariableTok{True}
\NormalTok{problem.model_selection.StabSel }\OperatorTok{=} \VariableTok{False}

\CommentTok{# Example how to add stability selection to the problem instance}
\NormalTok{problem.model_selection.StabSel }\OperatorTok{=} \VariableTok{True}
\end{Highlighting}
\end{Shaded}

Each model selection procedure has additional meta-parameters that are
described in the
\href{https://c-lasso.readthedocs.io/en/latest/}{Documentation}.

\section{Computational examples}\label{computational-examples}

\subsection{Toy example using synthetic
data}\label{toy-example-using-synthetic-data}

We illustrate the workflow of the \texttt{c-lasso} package on synthetic
data using the built-in routine \texttt{random\_data} which enables the
generation of test problem instances with normally distributed data
\(X\), sparse coefficient vectors \(\beta\), and constraints
\(C \in \mathbb{R}^{k\times d}\).

Here, we use a problem instance with \(n=100\), \(d=100\), a \(\beta\)
with five non-zero components, \(\sigma=0.5\), and a zero-sum contraint.

\begin{Shaded}
\begin{Highlighting}[]
\ImportTok{from}\NormalTok{ classo }\ImportTok{import}\NormalTok{ classo_problem, random_data}

\NormalTok{n,d,d_nonzero,k,sigma }\OperatorTok{=}\DecValTok{100}\NormalTok{,}\DecValTok{100}\NormalTok{,}\DecValTok{5}\NormalTok{,}\DecValTok{1}\NormalTok{,}\FloatTok{0.5}
\NormalTok{(X,C,y),sol }\OperatorTok{=}\NormalTok{ random_data(n,d,d_nonzero,k,sigma,zerosum}\OperatorTok{=}\VariableTok{True}\NormalTok{, seed }\OperatorTok{=} \DecValTok{123}\NormalTok{ )}
\BuiltInTok{print}\NormalTok{(}\StringTok{"Relevant variables  : }\SpecialCharTok{\{\}}\StringTok{"}\NormalTok{.}\BuiltInTok{format}\NormalTok{(}\BuiltInTok{list}\NormalTok{(numpy.nonzero(sol)) ) )}

\NormalTok{problem  }\OperatorTok{=}\NormalTok{ classo_problem(X,y,C)}

\NormalTok{problem.formulation.huber  }\OperatorTok{=} \VariableTok{True}
\NormalTok{problem.formulation.concomitant }\OperatorTok{=} \VariableTok{False}
\NormalTok{problem.formulation.rho }\OperatorTok{=} \FloatTok{1.5}

\NormalTok{problem.model_selection.LAMfixed }\OperatorTok{=} \VariableTok{True}
\NormalTok{problem.model_selection.PATH }\OperatorTok{=} \VariableTok{True}
\NormalTok{problem.model_selection.LAMfixedparameters.rescaled_lam }\OperatorTok{=} \VariableTok{True}
\NormalTok{problem.model_selection.LAMfixedparameters.lam }\OperatorTok{=} \FloatTok{0.1}

\NormalTok{problem.solve()}

\BuiltInTok{print}\NormalTok{(problem.solution)}
\end{Highlighting}
\end{Shaded}

We use \protect\hyperlink{formulations}{formulation}
\protect\hyperlink{R2}{\emph{R2}} with \(\rho=1.5\),
\protect\hyperlink{model}{computation mode and model selections}
\emph{Fixed Lambda} with \(\lambda = 0.1\lambda_{\max}\), \emph{Path
computation}, and \emph{Stability Selection} (as per default).

The corresponding output reads:

\begin{verbatim}
Relevant variables  : [43 47 74 79 84]

 LAMBDA FIXED : 
   Selected variables :  43    47    74    79    84    
   Running time :  0.294s

 PATH COMPUTATION : 
   Running time :  0.566s

 STABILITY SELECTION : 
   Selected variables :  43    47    74    79    84    
   Running time :  5.3s
\end{verbatim}

\texttt{c-lasso} allows standard visualization of the computed
solutions, e.g., coefficient plots at fixed \(\lambda\), the solution
path, the stability selection profile at the selected \(\lambda\), and
the stability selection profile across the entire path.

\begin{figure}
\centering
\includegraphics{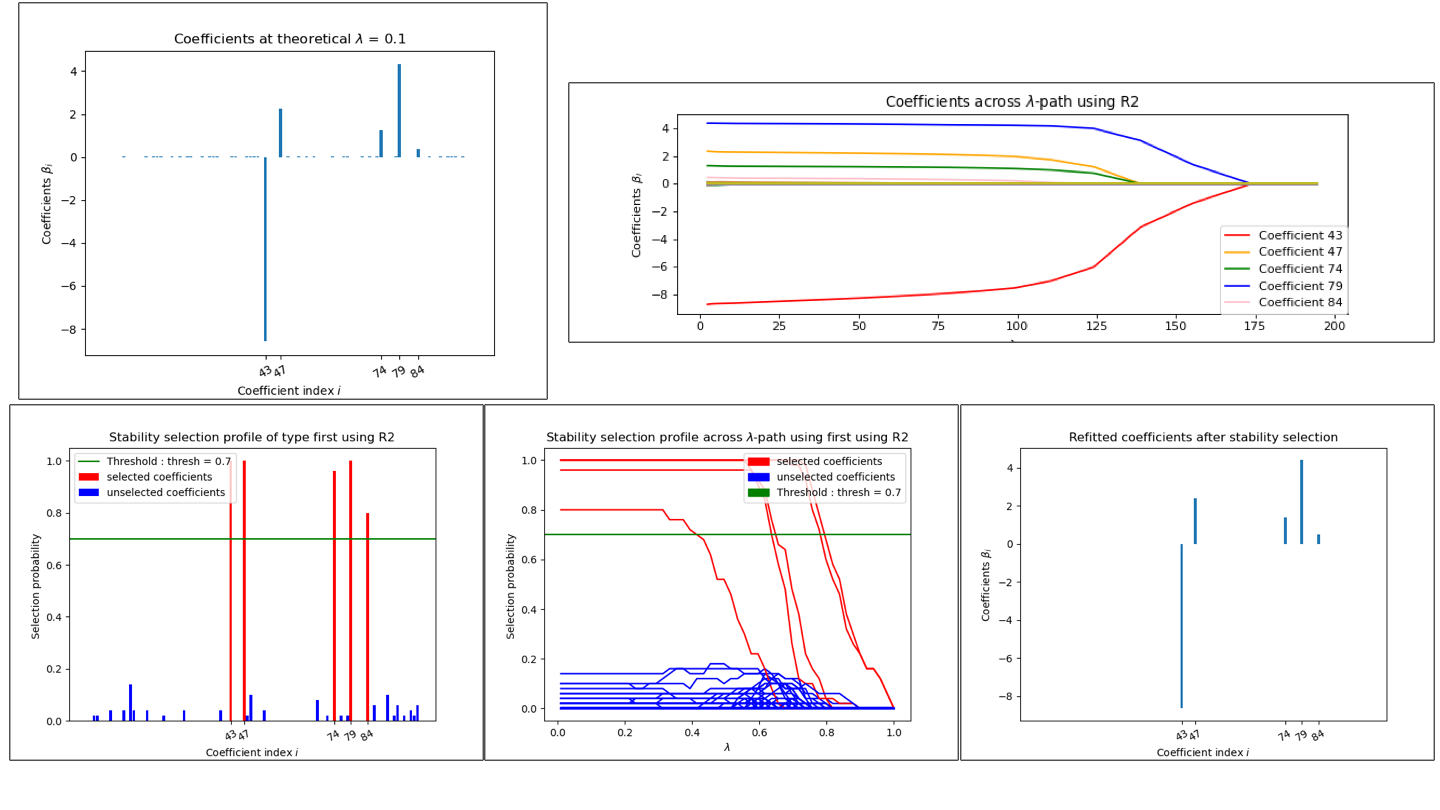}
\caption{Visualizations after calling problem.solution}
\end{figure}

For this tuned example, the solutions at the fixed lambda and with
stability selection recover the oracle solution. The solution vectors
are stored in \texttt{problem.solution} and can be directly acccessed
for each mode/model selection.

\begin{Shaded}
\begin{Highlighting}[]
\CommentTok{# Access to the estimated coefficient vector at a fixed lambda }
\NormalTok{problem.solution.LAMfixed.beta}
\end{Highlighting}
\end{Shaded}

Note that the run time for this \(d=100\)-dimensional example for a
single path computation is about 0.5 seconds on a standard Laptop.

\subsection{Log-contrast regression on gut microbiome
data}\label{log-contrast-regression-on-gut-microbiome-data}

We next illustrate the application of \texttt{c-lasso} on the
\href{https://github.com/Leo-Simpson/c-lasso/tree/master/examples/COMBO_data}{\texttt{COMBO}
microbiome dataset} (Lin et al. 2014; Shi, Zhang, and Li 2016; Combettes
and M\"uller 2020b). Here, the task is to predict the Body Mass Index
(BMI) of \(n=96\) participants from \(d=45\) relative abundances of
bacterial genera, and absolute calorie and fat intake measurements. The
code snippet for this example is available in the
\href{https://github.com/Leo-Simpson/c-lasso/README.md}{\texttt{README.md}}
and the
\href{https://github.com/Leo-Simpson/c-lasso/blob/master/examples/example-notebook.ipynb}{example
notebook}.

\begin{figure}
\centering
\includegraphics{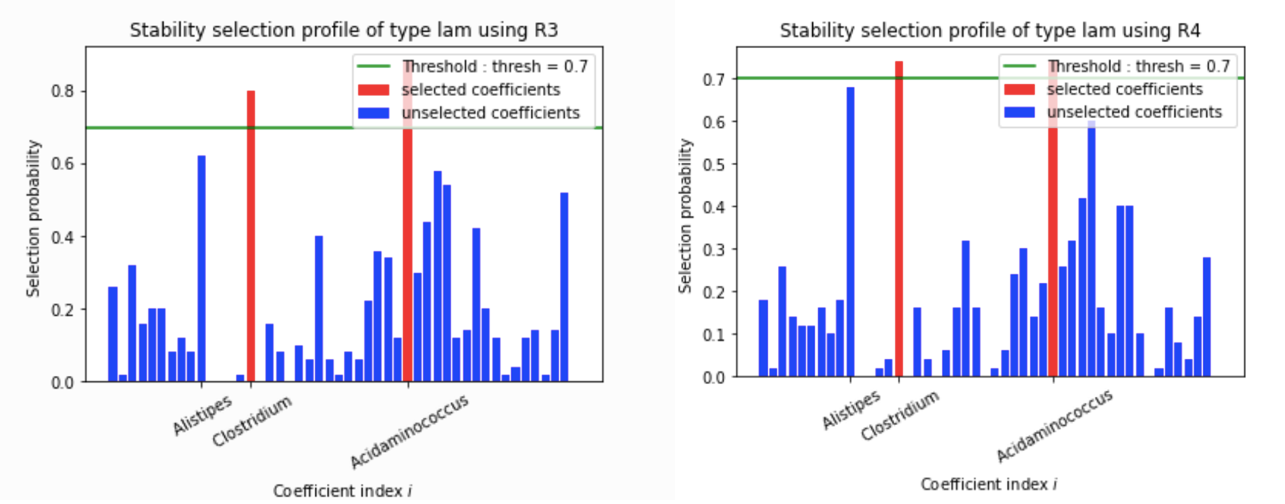}
\caption{Stability selection profiles of problems R3/R4 on the COMBO
data}
\end{figure}

Stability selection profiles using
\protect\hyperlink{formulations}{formulation}
\protect\hyperlink{R3}{\emph{R3}} (left) and
\protect\hyperlink{R4}{\emph{R4}}(right) on the COMBO dataset,
reproducing Figure 5a in (Combettes and M\"uller 2020b).

\subsection{\texorpdfstring{Calling \texttt{c-lasso} in
R}{Calling c-lasso in R}}\label{calling-c-lasso-in-r}

The \texttt{c-lasso} package also integrates with \texttt{R} via the
\texttt{R} package
\href{https://rstudio.github.io/reticulate/}{\texttt{reticulate}}. We
refer to \texttt{reticulate}'s manual for technical details about
connecting \texttt{python} environments and \texttt{R}. A successful use
case of \texttt{c-lasso} is available in the \texttt{R} package
\href{https://github.com/jacobbien/trac}{\texttt{trac}} (Bien et al.
2020), enabling tree-structured aggregation of predictors when features
are rare.

The code snippet below shows how \texttt{c-lasso} is called in
\texttt{R} to perform regression at a fixed \(\lambda\)
\(\lambda = 0.1\lambda_{\max}\). In \texttt{R}, X and C need to be of
\texttt{matrix} type, and y of \texttt{array} type.

\begin{Shaded}
\begin{Highlighting}[]
\NormalTok{problem <-}\StringTok{ }\NormalTok{classo}\OperatorTok{$}\KeywordTok{classo_problem}\NormalTok{(}\DataTypeTok{X=}\NormalTok{X,}\DataTypeTok{C=}\NormalTok{C,}\DataTypeTok{y=}\NormalTok{y) }
\NormalTok{problem}\OperatorTok{$}\NormalTok{model_selection}\OperatorTok{$}\NormalTok{LAMfixed <-}\StringTok{ }\OtherTok{TRUE}
\NormalTok{problem}\OperatorTok{$}\NormalTok{model_selection}\OperatorTok{$}\NormalTok{StabSel <-}\StringTok{ }\OtherTok{FALSE}
\NormalTok{problem}\OperatorTok{$}\NormalTok{model_selection}\OperatorTok{$}\NormalTok{LAMfixedparameters}\OperatorTok{$}\NormalTok{rescaled_lam <-}\StringTok{ }\OtherTok{TRUE}
\NormalTok{problem}\OperatorTok{$}\NormalTok{model_selection}\OperatorTok{$}\NormalTok{LAMfixedparameters}\OperatorTok{$}\NormalTok{lam <-}\StringTok{ }\FloatTok{0.1}
\NormalTok{problem}\OperatorTok{$}\KeywordTok{solve}\NormalTok{()}

\CommentTok{# Extract coefficent vector with tidy-verse}
\NormalTok{beta <-}\StringTok{ }\KeywordTok{as.matrix}\NormalTok{(}\KeywordTok{map_dfc}\NormalTok{(problem}\OperatorTok{$}\NormalTok{solution}\OperatorTok{$}\NormalTok{LAMfixed}\OperatorTok{$}\NormalTok{beta, as.numeric))}
\end{Highlighting}
\end{Shaded}

\section{Acknowledgements}\label{acknowledgements}

The work of LS was conducted at and financially supported by the Center
for Computational Mathematics (CCM), Flatiron Institute, New York, and
the Institute of Computational Biology, Helmholtz Zentrum M\"unchen. We
thank Dr.~Leslie Greengard (CCM and Courant Institute, NYU) for
facilitating the initial contact between LS and CLM. The work of PLC was
supported by the National Science Foundation under grant DMS-1818946.

\section*{References}\label{references}
\addcontentsline{toc}{section}{References}

\hypertarget{refs}{}
\hypertarget{ref-Aitchison:1984}{}
Aitchion, J., and J. Bacon-Shone. 1984. ``Log Contrast Models for
Experiments with Mixtures.'' \emph{Biometrika} 71 (2): 323--30.
doi:\href{https://doi.org/10.1093/biomet/71.2.323}{10.1093/biomet/71.2.323}.

\hypertarget{ref-Bien:2020}{}
Bien, Jacob, Xiaohan Yan, Léo Simpson, and Christian L M\"uller. 2020.
``Tree-Aggregated Predictive Modeling of Microbiome Data.''
\emph{bioRxiv}. Cold Spring Harbor Laboratory.
doi:\href{https://doi.org/10.1101/2020.09.01.277632}{10.1101/2020.09.01.277632}.

\hypertarget{ref-Briceno:2020}{}
Briceño-Arias, Luis, and Sergio López Rivera. 2019. ``A Projected
Primal--Dual Method for Solving Constrained Monotone Inclusions.''
\emph{Journal of Optimization Theory and Applications} 180 (3): 907--24.
doi:\href{https://doi.org/10.1007/s10957-018-1430-2}{10.1007/s10957-018-1430-2}.

\hypertarget{ref-Combettes:2020a}{}
Combettes, Patrick L., and Christian L. M\"uller. 2020a. ``Perspective
Maximum Likelihood-Type Estimation via Proximal Decomposition.''
\emph{Electron. J. Statist.} 14 (1). The Institute of Mathematical
Statistics; the Bernoulli Society: 207--38.
doi:\href{https://doi.org/10.1214/19-EJS1662}{10.1214/19-EJS1662}.

\hypertarget{ref-Combettes:2020b}{}
---------. 2020b. ``Regression Models for Compositional Data: General
Log-Contrast Formulations, Proximal Optimization, and Microbiome Data
Applications.'' \emph{Statistics in Biosciences}.
doi:\href{https://doi.org/10.1007/s12561-020-09283-2}{10.1007/s12561-020-09283-2}.

\hypertarget{ref-Combettes:2012}{}
Combettes, Patrick L., and Jean-Christophe Pesquet. 2012. ``Primal-Dual
Splitting Algorithm for Solving Inclusions with Mixtures of Composite,
Lipschitzian, and Parallel-Sum Type Monotone Operators.''
\emph{Set-Valued and Variational Analysis} 20 (June): 307--20.
doi:\href{https://doi.org/10.1007/s11228-011-0191-y}{10.1007/s11228-011-0191-y}.

\hypertarget{ref-Gaines:2018}{}
Gaines, Brian R., Juhyun Kim, and Hua Zhou. 2018. ``Algorithms for
Fitting the Constrained Lasso.'' \emph{Journal of Computational and
Graphical Statistics} 27 (4). Taylor \& Francis: 861--71.
doi:\href{https://doi.org/10.1080/10618600.2018.1473777}{10.1080/10618600.2018.1473777}.

\hypertarget{ref-Hastie:2009}{}
Hastie, T., R. Tibshirani, and J.H. Friedman. 2009. \emph{The Elements
of Statistical Learning: Data Mining, Inference, and Prediction}.
Springer Series in Statistics. Springer.
\url{https://books.google.fr/books?id=eBSgoAEACAAJ}.

\hypertarget{ref-Huber:1981}{}
Huber, P. 1981. \emph{Robust statistics}. John Wiley \& Sons Inc.

\hypertarget{ref-James:2020}{}
James, Gareth M., Courtney Paulson, and Paat Rusmevichientong. 2020.
``Penalized and Constrained Optimization: An Application to
High-Dimensional Website Advertising.'' \emph{Journal of the American
Statistical Association} 115 (529). Taylor \& Francis: 107--22.
doi:\href{https://doi.org/10.1080/01621459.2019.1609970}{10.1080/01621459.2019.1609970}.

\hypertarget{ref-Jeon:2020}{}
Jeon, Jong-June, Yongdai Kim, Sungho Won, and Hosik Choi. 2020. ``Primal
path algorithm for compositional data analysis.'' \emph{Computational
Statistics \& Data Analysis} 148: 106958.
doi:\href{https://doi.org/https://doi.org/10.1016/j.csda.2020.106958}{https://doi.org/10.1016/j.csda.2020.106958}.

\hypertarget{ref-Lee:2013}{}
Lee, Ching-Pei, and Chih-Jen Lin. 2013. ``A Study on L2-Loss (Squared
Hinge-Loss) Multiclass Svm.'' \emph{Neural Computation} 25 (March).
doi:\href{https://doi.org/10.1162/NECO_a_00434}{10.1162/NECO\_a\_00434}.

\hypertarget{ref-Lin:2014}{}
Lin, Wei, Pixu Shi, Rui Feng, and Hongzhe Li. 2014. ``Variable selection
in regression with compositional covariates.'' \emph{Biometrika} 101
(4): 785--97.
doi:\href{https://doi.org/10.1093/biomet/asu031}{10.1093/biomet/asu031}.

\hypertarget{ref-Meinshausen:2010}{}
Meinshausen, Nicolai, and Peter B\"uhlmann. 2010. ``Stability Selection.''
\emph{Journal of the Royal Statistical Society: Series B (Statistical
Methodology)} 72 (4): 417--73.
doi:\href{https://doi.org/10.1111/j.1467-9868.2010.00740.x}{10.1111/j.1467-9868.2010.00740.x}.

\hypertarget{ref-Mishra:2019}{}
Mishra, Aditya, and Christian L. M\"uller. 2019. ``Robust regression with
compositional covariates.'' \url{http://arxiv.org/abs/1909.04990}.

\hypertarget{ref-Rosset:2007}{}
Rosset, Saharon, and Ji Zhu. 2007. ``Piecewise linear regularized
solution paths.'' \emph{Annals of Statistics} 35 (3): 1012--30.
doi:\href{https://doi.org/10.1214/009053606000001370}{10.1214/009053606000001370}.

\hypertarget{ref-Shi:2016}{}
Shi, Pixu, Anru Zhang, and Hongzhe Li. 2016. ``Regression analysis for
microbiome compositional data.'' \emph{Annals of Applied Statistics} 10
(2): 1019--40.
doi:\href{https://doi.org/10.1214/16-AOAS928}{10.1214/16-AOAS928}.

\end{document}